\documentclass[amsmath,amssymb,aps,prd,showpacs,showkeys]{revtex4}
\usepackage{graphicx}
\usepackage{dcolumn}
\usepackage{bm}
\usepackage{latexsym,color}
\begin{document}
\def\be{\begin{equation}}
\def\ee{\end{equation}}
\def\bea{\begin{eqnarray}}
\def\eea{\end{eqnarray}}

\title{Hartle formalism for rotating Newtonian configurations} 

\author{Kuantay Boshkayev$^{1,2}$, Hernando Quevedo$^{2,3}$, Zhanerke Kalymova$^{1}$ and Bakytzhan Zhami$^{1}$}
\affiliation{
$^1$Faculty of Physics and Technology, Al-Farabi Kazakh National University, Al-Farabi av. 71, 050040 Almaty, Kazakhstan\\ 
$^2$Dipartimento di Fisica and ICRA, Universit\`a di Roma "La Sapienza",  Piazzale Aldo Moro 5, I-00185 Roma, Italy\\
$^3$Instituto de Ciencias Nucleares, Universidad Nacional Aut\'onoma de M\'exico, AP 70543, M\'exico, DF 04510, Mexico}
\email{kuantay@mail.ru,quevedo@nucleares.unam.mx}

\date{\today}

\begin{abstract}
We apply the Hartle formalism to study equilibrium configurations in the framework of Newtonian gravity. This approach allows one to study in a simple manner the properties of the interior gravitational field in the case of static as well as stationary rotating stars in hydrostatic equilibrium.  It is shown that the gravitational equilibrium conditions reduce to a system of ordinary differential equations which can be integrated numerically. We derive all the relevant equations up to the second order in the angular velocity. Moreover, we find explicitly the total mass, the moment of inertia, the quadrupole moment, the polar and equatorial radii, and the eccentricity of the rotating body. We also present the procedure to calculate the gravitational Love number. We test the formalism in the case of white dwarfs and show its compatibility with the known results in the literature.

\end{abstract}


\maketitle

\section{Introduction}
\label{intro}

In physics, rotation may introduce many changes in the structure of any system. In the case of celestial objects such as stars and planets,  rotation plays a crucial role. Rotation does not only change the shape of the celestial objects but also influences the processes occurring inside stars, i.e., it may accelerate or decelerate thermonuclear reactions under certain conditions, it changes the gravitational field outside the objects and it is one of the main factors that determines the lifespan of all stars (giant stars, main sequence, white dwarfs, neutron stars, etc.) \cite{Meynet2000,Meynet2003,Meynet2008,2013ApJ...762..117B,Boshkayev2014, boshmg13, belvedere2014}.

For instance, let us consider a white dwarf. A non-rotating white dwarf has a limiting mass of $ 1.44M_{\odot} $ which is well-known as the Chandrasekhar limit \cite{Chandrasekhar1931}. The central density and pressure corresponding to this limit define the evolution of white dwarfs. If the white dwarf rotates, then due to the centrifugal forces the central density and pressure decrease \cite{Chandrasekhar, shapirobook}. In order to recover the initial values of the central density and pressure of a rotating star one needs to add extra mass. Here we see that a rotating star with the same values for central density and pressure, as those of a non-rotating star, possesses a larger mass \cite{Stergioulas}.

In this work, we derive the equations describing the equilibrium configurations of slowly rotating stars within Hartle's formalism \cite{Hartle}. The advantage of this approach is that it allows us to consider in a simple way the influence of the rotation on the internal properties of the gravitational source. In fact, we will see that the complexity of the differential equations, which govern the dynamical properties of equilibrium configurations, is reduced to a high degree. When solving this kind of problems in celestial mechanics, astronomy and astrophysics, it is convenient to consider the internal structure of stars and planets as being described by a fluid. In the case of slow rotation, we derive equations that are valid  for any fluid up to the second order in the angular velocity.

As a result we obtain the equations defining the main parameters of the rotating equilibrium configuration such as the mass, radius, moment of inertia, gravitational potential, angular momentum and quadrupole moment as functions of the central density and angular velocity (rotation period). Furthermore, we show how to calculate the ellipticity and by means of it the gravitational Love number \cite{PoissonWill}. In turn, these parameters are of great importance in defining the evolution of a star.

In order to pursue all these issues in detail we revisit the Hartle formalism in classical physics for a slowly rotating configuration as the calculation of its equilibrium properties is much more simpler, because then the rotation can be considered as a small perturbation of an already-known non-rotating configuration. We therefore will consider in this paper a rotating configuration under the following conditions \cite{Hartle}:
\begin{itemize}
\item A one-parameter equation of state is specified, $p=p(\rho)$, where $p$ is the pressure and $\rho$ is the density of matter \cite{Chandrasekhar1939}.
\item A static equilibrium configuration is calculated using this equation of state and the classical equation of hydrostatic equilibrium for spherical symmetry.
\item Axial and reflection symmetry. The configuration is symmetric about a plane perpendicular to the axis of rotation.
\item A uniform angular velocity sufficiently slow so that the changes in pressure, energy density, and gravitational field are small.
\item Slow rotation. This requirement implies that the angular velocity $\Omega$ of the star
\begin{equation}\label{eq:1}
\Omega^2 \ll \frac{GM}{a^3},
\end{equation}
where $M$ is the mass of the unperturbed configuration, $a$ is its radius, $G$ is the gravitational constant. Consequently, the condition in Eq. (\ref{eq:1}) also implies
\begin{equation}\label{eq:2}
\Omega \ll \frac{c}{a} ,
\end{equation}
where $c$ is the speed of light.
\item The Newtonian field equations are expanded in powers of the angular velocity and the perturbations are calculated by retaining only the first- and second-order terms.
\end{itemize}
In the present work, the equations necessary to investigate  this issue are obtained explicitly. The problem of describing rotating configurations in Newtonian gravity has been investigated in many articles and textbooks \cite{Chandrasekhar1939, Jeffreys, Tassoul, chan33, psc71, jam64, ana65, rox65}. In this paper, we present in a pedagogical way a different approach. Indeed, we will use the Hartle formalism, which was originally proposed in general relativity, to consider a non-relativistic Newtonian configuration. The advantage of this method is that it can easily be included in an undergraduate course on astrophysics. No knowledge of relativistic physics is necessary, because only the very intuitive coordinate approach of Hartle's formalism is utilized in order to handle the Newtonian gravity equation and the corresponding equilibrium condition. It is worth mentioning that the Hartle formalism has been recently extended for relativistic configurations up to the fourth order terms in angular velocity \cite{yagi2014}.

This paper is organized as follows. In Sec. \ref{sec2}, we present the conditions under which the rotating compact object becomes a configuration in hydrostatic equilibrium. Moreover, we show that the use of a particular coordinate, which is especially adapted to describing the deformation due to the rotation, together with an expansion in spherical harmonics reduces substantially the system of differential equations up to the level that they can be integrated by quadratures. In Sec. \ref{sec:phys}, we derive expressions for the main physical quantities of the rotating object. A summary of the method to be used to find explicit numerical solutions by using our formalism is presented in Sec. \ref{sec:sum}. In Sec. \ref{sec:dwarfs} we apply the formalism to rotating white dwarfs in Newtonian physics and in Sec. \ref{kepl} we show the procedure of calculating the Keplerian angular velocity and the scaling law for the physical quantities describing rotating configurations. Finally, Sec. \ref{sec:con} is devoted to the conclusions and perspectives of our work. 
 

\section{Slowly rotating stars in Newtonian gravity}
\label{sec2}

In Newtonian gravitational theory the equilibrium configuration of uniformly rotating stars are determined by the solution of the three equations of Newtonian hydrostatic equilibrium \cite{Chandrasekhar, shapirobook, Hartle}. These are (1) the Newtonian field equation:
\begin{equation}\label{eq:3}
\nabla^2\Phi(r,\theta)=4\pi G\rho(r,\theta);
\end{equation}
where $\Phi$ is gravitational potential and  $\rho$ is the density of a fluid mass rotating with a uniform angular velocity $\Omega$;

(2) the equation of state that shows the relationship between pressure $p$ and density $\rho$ is assumed to have a one-parameter form
\begin{equation}\label{eq:4}
p=p(\rho);
\end{equation}

(3) the equation of hydrostatic equilibrium for uniformly rotating configurations which can be written as
\begin{equation}\label{eq:500}
\frac{d\vec{v}}{dt}=-\frac{1}{\rho}\vec{\nabla} p-\vec{\nabla} \Phi \ ,
\end{equation}
with
\begin{equation}\label{eq:5000}
\vec{v}=\frac{d\vec{r}}{dt}=\vec{\Omega}\times\vec{r} \ .
\end{equation}
For uniform rotation ($\Omega=const$) we have that
\begin{equation}\label{eq:50000}
\frac{d\vec{v}}{dt}=\vec{\Omega}\times\vec{v}=\vec{\Omega}\times(\vec{\Omega}\times\vec{r})=-\frac{1}{2}\vec{\nabla}(\vec{\Omega}\times\vec{r})^2 \ .
\end{equation}
Therefore, substituting this expression in (\ref{eq:500}), we obtain
\begin{equation}\label{eq:500000}
-\frac{1}{2}\vec{\nabla}(\vec{\Omega}\times\vec{r})^2=-\frac{1}{\rho}\vec{\nabla} p-\vec{\nabla} \Phi
\end{equation}
or
\begin{equation}\label{eq:5000000}
\frac{dp}{\rho}-\frac{1}{2}d(\vec{\Omega}\times\vec{r})^2+d\Phi=0 \ ,
\end{equation}
which can be reexpressed in terms of its first integral
\begin{equation}\label{eq:5}
\int_{0}^{p}\frac{dp(r,\theta)}{\rho(r,\theta)}-\frac{1}{2}\Omega^2r^2\sin^2\theta+\Phi(r,\theta)=const,
\end{equation}
where $r$ is the radial coordinate and $\theta$ is the polar coordinate of the rotating configuration.

The main task now is to expand the equations of Newtonian hydrostatics  in powers of $\Omega^2$. The solution to the first term of the expansion is given by $\Phi^{(0)}$, $p^{(0)}$, and $\rho^{(0)}$ in the absence of rotation. Then, it is necessary to find the equations which govern the second-order terms. It is expected that the resulting differential equations can be integrated in terms of the known non-rotating solution.


\subsection{{Coordinates}}
\label{sec3}

\begin{figure}
\includegraphics[width=3in]{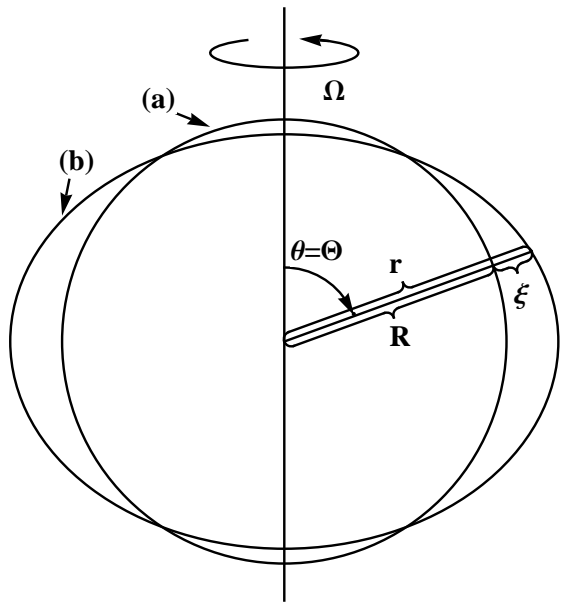} 
\caption{{\footnotesize Definition of the coordinates $R$, $\Theta$, and the displacement $\xi(R,\Theta)$. The surface (a) is the surface of constant density $\rho(R)$ in the non-rotating configuration. The surface (b) is the surface of constant density $\rho(R)$ in the rotating configuration (reproduced from \cite{Hartle}). }}\label{fig:1} 
\end{figure}

An important point to be considered is the choice of the coordinate system in which the expansions in powers of $\Omega$ are carried out. As pointed out by Hartle in 1967 \cite{Hartle}, one should be very careful when considering perturbation near the surface of the star. Therefore, we select a coordinate transformation such that the density of the star in terms of the new radial coordinate is the same as in the static configuration: 
\begin{equation}\label{eq:7}
\rho[r(R,\Theta),\Theta]=\rho(R)=\rho^{(0)}(R).
\end{equation}
Thus, the  relationships between the old coordinates ($r, \theta$) and the new coordinates ($R,\Theta$) are given by
\begin{eqnarray}\label{eq:6}
\theta&=&\Theta, \qquad r=R+\xi(R,\Theta)+O(\Omega^4)\ . 
\end{eqnarray}
The function $r(R,\Theta)$ then replaces the density as a function to be calculated in the rotating configuration. These definitions are given pictorially in Fig. \ref{fig:1}.

Following Hartle's formalism, we are always free to consider the rotating configuration as a perturbation of a non-rotating configuration with the same central density. Consequently, in the $R,\Theta$ coordinate system, the density (\ref{eq:7}) and pressure are known functions of $R$ 
\begin{eqnarray}\label{eq:9}
p[r(R,\Theta),\Theta]=p(R)=p^{(0)}(R) 
\end{eqnarray}
related by the one-parametric equation of state. 


\subsection{Spherical harmonics}

The expansion of $r$ in terms of $\Omega^2$ is given by equation (\ref{eq:6}) and the expansion of the gravitational potential $\Phi$ can be represented as 
\begin{eqnarray}\label{eq:10}
\Phi(R,\Theta)&\approx&\Phi^{(0)}(R)+\Phi^{(2)}(R,\Theta)+O(\Omega^4)
\end{eqnarray}
where $\Phi^{(0)}(R)$ is the spherical part of the potential and $\Phi^{(2)}(R,\Theta)$ is the perturbed part. Calculating the Taylor expansion in terms of the new the coordinates, we obtain
\begin{eqnarray}\label{eq:1000}
\Phi(r,\theta)&=&\Phi(R+\xi,\Theta)\approx\Phi(R,\Theta)+\xi\frac{d\Phi(R,\Theta)}{dR}+O(\Omega^4) \\ \nonumber 
&\approx&\Phi^{(0)}(R)+\xi\frac{d\Phi^{(0)}(R)}{dR}+\Phi^{(2)}(R,\Theta)+O(\Omega^4).
\end{eqnarray}


In order to simplify the equations we expand the functions $\xi$ and $\Phi^{(2)}$ in spherical harmonics
\begin{eqnarray}\label{eq:11}
\xi(R,\Theta)&=&\sum_{l=0}^{\infty}{\xi_{l}(R)P_{l}(\cos\Theta)},\quad \Phi^{(2)}(R,\Theta)=\sum_{l=0}^{\infty}{\Phi^{(2)}_{l}(R)P_{l}(\cos\Theta)},\quad \Phi^{(2)}(R,\Theta)\sim\Omega^2,
\end{eqnarray}
where $P_{l}(\cos\Theta)$ are the Legendre polynomials. 



Now let us perform the computations in detail taking the polar axis to be the axis of rotation. Using the expressions for the Legendre polynomials $P_{0}(\cos\Theta)=1$ and 
$P_{2}(\cos\Theta)=\frac{1}{2}(3\cos^2\Theta-1)$, it is easy to show that 
\begin{equation}
\sin^2\Theta=\frac{2}{3}[P_{0}(\cos\Theta)-P_{2}(\cos\Theta)],
\end{equation}
From here we see that $l$ accepts only two values, namely, $0$ and $2$. The equations for $\xi_{l}(R)$, $\Phi_{l}^{(2)}(R)$, with $l\geq4$ are thus independent of $\Omega$ and their solution is

\begin{equation}\label{eq:12}
\xi_{l}=0, \qquad \Phi_{l}^{(2)}=0, \qquad l\geq4.
\end{equation}

Rewriting the condition  of hydrostatic equilibrium (\ref{eq:5})  in coordinates $(R, \Theta)$ and expanding it in spherical harmonics by using  Eqs. (\ref{eq:7}), (\ref{eq:10}), and (\ref{eq:11}), we get
\begin{eqnarray}
\int_{0}^{p}\frac{dp^{(0)}(R)}{\rho(R)}-\frac{1}{3}\Omega^2R^2[P_{0}(\cos\Theta)-P_{2}(\cos\Theta)]+\Phi^{(0)}(R)\qquad\qquad\\ \nonumber
+\sum_{l=0}^{n}{\Phi^{(2)}_{l}(R)P_{l}(\cos\Theta)}+\sum_{l=0}^{n}{\xi_{l}(R)P_{l}(\cos\Theta)}\frac{d\Phi^{(0)}(R)}{dR}=const\ .
\end{eqnarray}
We now collect the terms proportional to  $\sim\Omega^0$ and $\Omega^2$ with $l=0,2$ and obtain:
\begin{equation}\label{eq:16}
\int_{0}^{p}\frac{dp^{(0)}(R)}{\rho(R)}+\Phi^{(0)}(R)=const,
\end{equation}
\begin{eqnarray}
-\frac{1}{3}\Omega^2R^2+\Phi^{(2)}_{0}(R)+\xi_{0}(R)\frac{d\Phi^{(0)}(R)}{dR}&=&0\label{eq:17},\\
\frac{1}{3}\Omega^2R^2+\Phi^{(2)}_{2}(R)+\xi_{2}(R)\frac{d\Phi^{(0)}(R)}{dR}&=&0\label{eq:18}.
\end{eqnarray}
The first of the above equations corresponds to the Newtonian hydrostatic equation for a static configuration.

Using the same procedure, the Newtonian field equation becomes
\begin{eqnarray}
\nabla^2\Phi(r,\theta) =\frac{1}{r^2}\frac{\partial}{\partial r}\left(r^2\frac{\partial \Phi(r,\theta)}{\partial r}\right) +\frac{1}{r^2\sin\theta}\frac{\partial}{\partial \theta}\left(\sin\theta\frac{\partial \Phi(r,\theta)}{\partial\theta}\right) &&\\ \nonumber
=\nabla^2_{r}\Phi(r,\theta)+\frac{1}{r^2}\nabla^2_{\theta}\Phi(r,\theta)\approx\nabla^2_{r}\Phi^{(0)}(r)+\nabla^2_{r}\Phi^{(2)}_{0}(r) &&\\ \nonumber +\nabla^2_{r}\Phi^{(2)}_{2}(r)P_{2}(\cos\theta)+\frac{1}{r^2}\nabla^2_{\theta}\Phi^{(2)}_{2}(r)P_{2}(\cos\theta)=4\pi G\rho(r,\theta). &&
\end{eqnarray}
Since the functions $\Phi_0^{(2)}$ and $\Phi_2^{(2)}$ are already proportional to $\Omega^2$, we can directly write them in $(R,\Theta)$ coordinates. 
However $\nabla^2_{r}\Phi^{(0)}(r)\approx\nabla^2_{R}\Phi^{(0)}(R)+\xi(R,\Theta)\frac{d}{dR}\nabla^2_{R}\Phi^{(0)}(R)$ (see the Appendix for more details). Thus
\begin{eqnarray}
&&\nabla^2\Phi(r,\theta)=\nabla^2_{R}\Phi^{(0)}(R)+\xi(R,\Theta)\frac{d}{dR}\nabla^2_{R}\Phi^{(0)}(R) \\  \nonumber &&+\nabla^2_{R}\Phi^{(2)}_{0}(R)+\nabla^2_{R}\Phi^{(2)}_{2}(R)P_{2}(\cos\Theta)+\frac{1}{R^2}\nabla^2_{\Theta}\Phi^{(2)}_{2}(R)P_{2}(\cos\Theta)=4\pi G\rho(R).
\end{eqnarray}
Taking into account that $\xi(R,\Theta)=\xi_0(R)+\xi_2(R)P_2(\cos\Theta)$ and collecting the corresponding terms, we obtain  the Newtonian field equations of both static and rotating configurations:
\begin{eqnarray}
&&\nabla^2_{R}\Phi^{(0)}(R)=4\pi G\rho(R),\\ 
&&\xi_{0}(R)\frac{d}{dR}\nabla^2_{R}\Phi^{(0)}(R)+\nabla^2_{R}\Phi^{(2)}_{0}(R)=0 \label{eq:22},\\  
&&\xi_{2}(R)\frac{d}{dR}\nabla^2_{R}\Phi^{(0)}(R)+\nabla^2_{R}\Phi^{(2)}_{2}(R)-\frac{6}{R^2}\Phi^{(2)}_{2}(R)=0 \label{eq:23}. 
\end{eqnarray}

The differential equations for $\Phi_0^{(2)}(R)$, $\Phi_2^{(2)}(R)$, $\xi_0(R)$, and $\xi_2(R)$, which establish the relation between mass and central density for rotating star and determine the shape of the star, will now be given in forms suitable for solving these problems.

\section{Physical properties of the model}
\label{sec:phys}

The above description of the rotating equilibrium configuration allows us to derive all the main quantities that are necessary for establishing  the physical significance and determining the physical properties of the rotating source. In this section, we will derive all the equations that must be solved in order to find the values of all the relevant quantities.

\subsection{Mass and Central Density}

The total mass of the rotating configuration is given by the integral of the density over the volume,
\be
M_{tot} =\int_{V}\rho(r,\theta)dV=\int_{V}\rho(r,\theta)r^2dr\sin\theta d\theta d\phi \ .\nonumber
\ee
To proceed with the computation of the integral, we use formula (\ref{eq:6}) and obtain the relationship
\be
r^2dr=(R+\xi)^2(dR+d\xi)\approx R^2\left(1+\frac{2\xi}{R}\right)\left(1+\frac{d\xi}{dR}\right)dR=\left(1+\frac{2\xi}{R}+\frac{d\xi}{dR}\right)R^2dR\ ,
\ee
which implies that 
\begin{eqnarray}
M_{tot} &=&\int_{V}\rho(R)R^2dR\sin\Theta d\Theta d\phi  \\ \nonumber
&+&\int_{V}\rho(R)R^2\left(\frac{2\xi(R,\Theta)}{R}+\frac{d\xi(R,\Theta)}{dR}\right)dR\sin\Theta d\Theta d\phi \ .
\end{eqnarray}

Performing the integration within the range of angles $0<\Theta<\pi$ and $0<\phi<2\pi$ and using the identities
\begin{eqnarray}
\int_{0}^{\pi}\sin\Theta d\Theta&=&2,\\
\int_{0}^{\pi}P_2(\cos\Theta)\sin\Theta d\Theta&=&0 \ ,
\end{eqnarray}
one finds that the change in mass $M^{(2)}$ of the rotating configuration from the non-rotating one can be written as
\begin{eqnarray}
M_{tot}(R)&=&M^{(0)}(R)+M^{(2)}(R), \label{eq:25}\\ 
M^{(0)}(R)&=&4\pi\int_{0}^{R} \rho(R)R^2 dR,\\ 
M^{(2)}(R)&=&4\pi\int_{0}^{R}\rho(R)R^2\left(\frac{2\xi_0(R)}{R}+\frac{d\xi_0(R)}{dR}\right)dR =4\pi \int_{0}^{R}\left(-\xi_{0}(R)\frac{d\rho(R)}{dR}\right)R^2dR\ .
\end{eqnarray}
Here we have used the following expressions that follow from the field equations and definitions of the masses
\begin{eqnarray}
\nabla^2\Phi^{(0)}(R)&=&4\pi G \rho(R), \\
\frac{d}{dR}\nabla^2\Phi^{(0)}(R)&=&4\pi G \frac{d\rho(R)}{dR}, \\
\frac{dM^{(0)}(R)}{dR}&=&4\pi R^2 \rho(R),\\
\frac{dM^{(2)}(R)}{dR}&=&4\pi \left(-\xi_{0}(R)\frac{d\rho(R)}{dR}\right)R^2 \ .
\end{eqnarray}

Using the condition that $\Phi^{(0)}(R), \Phi_{0}^{(2)}(R)\rightarrow const$, as $R\rightarrow 0$, and taking into account (\ref{eq:22}) the masses of both configurations can be expressed as
\begin{eqnarray}
\frac{GM^{(0)}(R)}{R^2}=\frac{d\Phi^{(0)}(R)}{dR},\\
\frac{GM^{(2)}(R)}{R^2}=\frac{d\Phi_{0}^{(2)}(R)}{dR}.\label{eq:37000}
\end{eqnarray}

It is convenient to display the $l=0$ equation in a form in which it resembles the equation of hydrostatic equilibrium. To do this, we define
\begin{eqnarray}
p_0^*(R)&=&\xi_{0}(R)\frac{d\Phi^{(0)}(R)}{dR}\ .
\end{eqnarray}
Moreover, taking derivative of (\ref{eq:17}) and taking into account (\ref{eq:37000}), we obtain
\begin{eqnarray}\label{eq:37}
-\frac{dp_0^*(R)}{dR}+\frac{2}{3}\Omega^2R=\frac{GM^{(2)}(R)}{R^2} \ .
\end{eqnarray}
The above equation along with
\begin{eqnarray}\label{eq:38}
\frac{dM^{(2)}(R)}{dR}=4\pi R^2\rho(R)\frac{d\rho(R)}{dp(R)}p_0^*(R),
\end{eqnarray}
show the balance between the pressure, centrifugal, and gravitational forces per unit mass in the rotating star. 
The latter expression was obtained by using (\ref{eq:16}).

%
%
%


\subsection{The Shape of the Star and Numerical Integration}

If the surface of the non-rotating star has radius $a$, then equations (\ref{eq:7}) and (\ref{eq:11}) show that the equation for the surface of the rotating star has the form
\begin{eqnarray}\label{eq:40}
r(a,\Theta)&=&a+\xi_{0}(a)+\xi_{2}(a)P_2(\cos\Theta).
\end{eqnarray}\label{eq:41}
The value of $\xi_{0}(a)$ is already determined in the $l=0$ calculation
\begin{eqnarray}
\xi_{0}(a)&=&\frac{a^2}{GM}p_0^*(a),
\end{eqnarray}
where $M=M^{(0)}(a)$ is the mass of the non-rotating configuration.
However, the determination of $\xi_2(R)$ from $l=2$ equations is not straightforward. So far, we have the $l=2$ equations (\ref{eq:18}) and (\ref{eq:23}) representing the hydrostatic equilibrium and the field equation, respectively. From (\ref{eq:18}) we obtain the expression 
\be
\xi_{2}(R)=-\frac{R^2}{GM(R)}\left\{\frac{1}{3}\Omega^2R^2+\Phi^{(2)}_{2}(R)\right\},\label{eq:42}
\ee
which we insert into $(\ref{eq:23})$ and get
\be
\nabla^2_{R}\Phi^{(2)}_{2}(R)-\frac{6}{R^2}\Phi^{(2)}_{2}(R)=\frac{4\pi R^2}{M(R)}\left\{\frac{1}{3}\Omega^2R^2+\Phi^{(2)}_{2}(R)\right\}\frac{d\rho(R)}{dR},
\label{eq:43} 
\ee
where $M(R)=M^{(0)}(R)$ denotes the non-rotating mass. 
In order to solve the latter equation numerically, one needs to rewrite it as first-order linear differential equations. To this end, we introduce 
new functions $\varphi=\Phi^{(2)}_{2}$ and $\chi$ so that Eq. (\ref{eq:43}) generates the system 
\begin{eqnarray}
&&\frac{d\chi(R)}{dR}=-\frac{2GM(R)}{R^2}\varphi(R)+\frac{8\pi}{3}\Omega^2R^3 G \rho(R),\label{eq:44}\\
&&\frac{d\varphi(R)}{dR}=\left(\frac{4\pi R^2\rho(R)}{M(R)}-\frac{2}{R}\right)\varphi(R)-\frac{2\chi(R)}{GM(R)}+\frac{4\pi}{3M(R)}\rho(R) \Omega^2 R^4. \label{eq:45}
\end{eqnarray}

The above equations can be solved by quadratures. The computation of the solution can be performed numerically by integrating outward from the origin. 
At the origin the solution must be regular. An examination of the equations shows that, as $R\rightarrow0$,
\begin{eqnarray}
\varphi(R)&\rightarrow&A R^2, \qquad \chi(R)\rightarrow B R^4,
\end{eqnarray}
where $A$ and $B$ are any constants related by
\begin{eqnarray}\label{AB}
B+\frac{2\pi}{3}G\rho_{c} A=\frac{2\pi}{3}G\rho_{c}\Omega^2
\end{eqnarray}
and  $\rho_{c}$ is the value of the density in the center of the star.
The remaining constant in the solution is determined by the boundary condition that $\varphi(R)\rightarrow0$ at large values of $R$. The constant is thus determined by matching the interior solution with the exterior solution which satisfies this boundary condition.

In the exterior region, the solutions of the equations (\ref{eq:44}) and (\ref{eq:45}) are
\begin{eqnarray}\label{ExterSol}
\varphi_{ex}(R)&=&\frac{K_1}{R^3},\qquad \chi_{ex}(R)=\frac{K_1 G M^{(0)}}{2R^4}.
\end{eqnarray}
The interior solution to the equations  (\ref{eq:44}) and (\ref{eq:45}) may be written as the sum  of a particular solution and a homogeneous solution. The particular solution may be obtained by integrating the equations outward from the center with any values of $A$ and $B$ which satisfy (\ref{AB}). The homogeneous solution is then obtained by integrating the equations
\begin{eqnarray}
&&\frac{d\chi_h(R)}{dR}=-\frac{2GM(R)}{R^2}\varphi_h(R),\\
&&\frac{d\varphi_h(R)}{dR}=\left(\frac{4\pi R^2\rho(R)}{M(R)}-\frac{2}{R}\right)\varphi_h(R)-\frac{2\chi_h(R)}{GM(R)},
\end{eqnarray}
with $A$ and $B$ related now by
\begin{eqnarray}
B+\frac{2\pi}{3}G\rho_{c}A=0
\end{eqnarray}

The general solution may then be written as
\begin{eqnarray}\label{InterSol}
\varphi_{in}(R)=\varphi_{p}(R)+K_2\varphi_{h}(R) ,\qquad \chi_{in}(R)=\chi_{p}(R)+K_2\chi_{h}(R).
\end{eqnarray}
By matching (\ref{ExterSol}) and (\ref{InterSol}) at $R=a$, the constants $K_1$ and $K_2$ can be determined. Thus, $\varphi_{in}(R)$ is determined and $\xi_2(R)$ can be easily calculated from
\begin{eqnarray}\label{eq:53}
&&\xi_{2}(R)=-\frac{R^2}{GM(R)}\left\{\frac{1}{3}\Omega^2R^2+\varphi_{in}(R)\right\}.
\end{eqnarray}


\subsection{Moment of Inertia}

Similarly to the total mass of the star, the total moment of inertia can be calculated as
\be
I_{tot} =\int_{V}\rho(r,\theta)(r\sin\theta)^2dV=\int_{V}\rho(r,\theta)r^4dr\sin^3\theta d\theta d\phi \ .
\label{min}
\ee
Using the definition of the radial coordinate $r$, we find the expression
\be
r^4dr=(R+\xi)^4(dR+d\xi)\approx R^4\left(1+\frac{4\xi}{R}\right)\left(1+\frac{d\xi}{dR}\right)dR=\left(1+\frac{4\xi}{R}+\frac{d\xi}{dR}\right)R^4dR \ ,
\ee
which allows us to rewrite the moment of inertia as 
\begin{eqnarray}
I_{tot} &=&\int_{V}\rho(R)R^4dR\sin^3\Theta d\Theta d\phi  \\ \nonumber
&+&\int_{V}\rho(R)R^4\left(\frac{4\xi(R,\Theta)}{R}+\frac{d\xi(R,\Theta)}{dR}\right)dR\sin^3\Theta \nonumber d\Theta d\phi\ .
\end{eqnarray}

Performing the integration within the range  $0<\Theta<\pi$ and $0<\phi<2\pi$, we obtain
\begin{eqnarray}
I_{tot}(R)&=&I^{(0)}(R)+I^{(2)}(R), \\ \nonumber
I^{(0)}(R)&=&\frac{8\pi}{3}\int_{0}^{R} \rho(R)R^4 dR ,\\ \nonumber
I^{(2)}(R)&=&\frac{8\pi}{3}\int_{0}^{R}\rho(R)R^4 \left(\frac{d\xi_0(R)}{dR}-\frac{1}{5}\frac{d\xi_2(R)}{dR}+\frac{4}{R}\left[\xi_0(R)-\frac{1}{5}\xi_2(R)\right]\right) dR \\ \nonumber
&=& \frac{8\pi}{3}\int_{0}^{R}\left(\left[\frac{1}{5}\xi_2(R)-\xi_0(R)\right]\frac{d\rho(R)}{dR}\right)R^4  dR,
\end{eqnarray}
where we have used the integrals 
\begin{eqnarray}
\int_{0}^{\pi}\sin^3\Theta d\Theta&=&\frac{4}{3},\\
\int_{0}^{\pi}P_2(\cos\Theta)\sin^3\Theta d\Theta&=&-\frac{4}{15}.
\end{eqnarray}

In the corresponding limit, our results coincide with the definition of the moment of inertia for slowly rotating relativistic stars as given in  \cite{Hartle2}. Notice that, knowing the value of the moment of inertia, one can easily calculate the total angular momentum of the rotating stars
\begin{eqnarray}
J_{tot}=J^{(0)}+J^{(2)},
\end{eqnarray}
where $J^{(0)}=I^{(0)}\Omega$ is the angular momentum of the spherical configuration and $J^{(2)}=I^{(2)}\Omega$ is the change of the angular momentum due to rotation and deformation.


\subsection{Quadrupole Moment}

The Newtonian potential $\Phi(R,\Theta)$ outside the star will be written as before as (see Eq.(\ref{eq:10}))
\begin{eqnarray}\label{eq:57}
&&\Phi(R,\Theta)=\Phi^{(0)}(R)+\Phi^{(2)}_0(R)+\Phi^{(2)}_2(R)P_2(\cos\Theta),
\end{eqnarray}
where
\begin{eqnarray}
\Phi^{(0)}(R)&=&-\frac{GM^{(0)}}{R}, \\
\Phi^{(2)}_0(R)&=&-\frac{GM^{(2)}}{R}, \\
\Phi^{(2)}_2(R)&=&\frac{K_1}{R^3}.
\end{eqnarray}
In view of (\ref{eq:25}), equation (\ref{eq:57}) can be written as follows
\begin{eqnarray}
&&\Phi(R,\Theta)=-\frac{GM_{tot}}{R}+\frac{K_1}{R^3}P_2(\cos\Theta),
\end{eqnarray}

It follows that the constant $K_1$ determines the mass quadrupole moment $Q$ of the star as $K_1=G Q$. For a vanishing $K_1$ we recover the non-rotating configuration. Moreover, according to Hartle's definition $Q>0$ represents an oblate object and $Q<0$ corresponds to a prolate object.


\subsection{Ellipticity and Gravitational Love Number}

The quantity defined by 
\begin{equation}\label{eq:56el0}
\epsilon(R)=-\frac{3}{2R}\xi_{2}(R),
\end{equation}
is the ellipticity of the surface of constant density labeled by $R$. We use this expression and (\ref{eq:42}), and eliminate $\Phi_2^{(2)}$ from (\ref{eq:43}), to obtain the following equation for $\epsilon(R)$:
\begin{equation}\label{eq:56el}
\frac{M(R)}{R}\frac{d^2 \epsilon(R)}{dR^2}+\frac{2}{R}\frac{dM(R)}{dR}\frac{d\epsilon(R)}{dR}+\frac{2dM(R)}{dR}\frac{\epsilon(R)}{R^2}-\frac{6M(R)\epsilon(R)}{R^3}=0,
\end{equation}
or equivalently in a compact form 
\begin{equation}\label{eq:56}
\frac{d}{dR}\frac{1}{R^4}\frac{d}{dR}\left[\epsilon(R)M(R)R^2\right]=4\pi\epsilon(R)\frac{d\rho(R)}{dR} \ .
\end{equation}
This equation is equivalent to Clairaut's equation. Here both $M(R)$ and $\rho(R)$ are known functions of $R$. The ellipticity must be regular at small values of $R$, and equation (\ref{eq:56}) shows that it approaches a constant at $R=0$. With this boundary condition,  equation (\ref{eq:56}) may be integrated to find the shape of $\epsilon(R)$. To find the magnitude of $\epsilon(R)$ one needs to use (\ref{eq:53}). The procedure for considering the boundary condition at the surface  given in the previous section, together with the condition of regularity at the origin and the differential equation (\ref{eq:56}), uniquely determine  the ellipticity of the surfaces of constant density as a function of the coordinate $R$. 

It is easy to show that equation (\ref{eq:56el}) can be written in the form given in  Ref. \cite{Tassoul}
\begin{equation}\label{eq:56el2}
R^2\frac{d^2 \epsilon(R)}{dR^2}+6\frac{\rho(R)}{\rho_{m}(R)}\left[R\frac{d\epsilon(R)}{dR}+\epsilon(R)\right]=6\epsilon(R),
\end{equation}
where
\begin{equation}\label{eq:56el3}
\rho_{m}(R)=\frac{4\pi R^3}{3M(R)}
\end{equation}
is the average mass density.
By introducing a new function as 

\begin{equation}\label{eq:56el4}
\eta_{2}(R)=\frac{R}{\epsilon(R)}\frac{d\epsilon(R)}{dR} \ ,
\end{equation}
Eq.(\ref{eq:56el2}) reduces to the well known  Clairaut-Radau equation \cite{PoissonWill}
\begin{equation}\label{eq:56el5}
R\frac{d \eta_{2}(R)}{dR}+6\mathcal D(R)[\eta_{2}(R)+1]+\eta_{2}(R)[\eta_{2}(R)-1]=6,
\end{equation}

where

\begin{equation}\label{eq:56el6}
\mathcal D(R)=\frac{\rho(R)}{\rho_{m}(R)}
\end{equation}
encodes the relevant information about the structure of the body. The differential equation is integrated outward from $R=0$, with the boundary conditions $\mathcal D(0)=1$ and $\eta(R=0)=0$, up to $R=a$, obtaining the value $\eta_{2}(R=a)$.  The Love number is then given by 

\begin{equation}\label{eq:56el7}
k_2=\frac{3-\eta_{2}(a)}{2[2+\eta_{2}(a)]}
\end{equation}

Note, once $\xi_{2}(R)$ is known then $\epsilon(R)$ is also known from Eq. (\ref{eq:56el0}) and we have

\begin{equation}\label{eq:56el8}
\eta_2(a)=\frac{a}{\epsilon(a)}\frac{d\epsilon(R)}{dR}|_{R=a}=\frac{a}{\xi_{2}(a)}\frac{d\xi_{2}(R)}{dR}|_{R=a}-1
\end{equation}

One can see from here that $\eta_2(a)$ does not depend on the angular velocity of the star, neither does the Love number.


\section{Summary}
\label{sec:sum}

Our results show that it is possible to write explicitly all the differential equations that determine the behavior of a slowly rotating compact object. For a better presentation of the results obtained in preceding sections, we summarize the steps that must be followed to integrate the resulting equations. 

\subsection{The static case}

To determine the relation between mass and central density,  one must proceed as follows. 
(1) Specify the equation of state $p=p(\rho)$ (polytrope, tabulated, etc.). 
(2) Choose the value of the central density $\rho(R=0)=\rho_c$. Calculate the mass and pressure from the Newtonian field equation and the equation of hydrostatic equilibrium with the regularity condition at the center $M^{(0)}(R=0)=0$
\begin{equation}
\begin{cases}
\frac{dM^{(0)}(R)}{dR}=4\pi R^2 \rho(R),\\
\frac{dp^{(0)}(R)}{dR}=-\rho(R)\frac{GM^{(0)}(R)}{R^2} .
\end{cases}
\end{equation}
The gravitational potential of the  non-rotating star is obtained as
\begin{equation}
\frac{d\Phi^{(0)}(R)}{dR}=\frac{GM^{(0)}(R)}{R^2}=-\frac{1}{\rho(R)}\frac{dp^{(0)}(R)}{dR}\ .
\end{equation}
On the surface, the pressure must vanish $p^{(0)}(R=a)=0$.

\subsection{The rotating case: $l=0$ Equations}

Select the value of the angular velocity of the star. For instance, take as a test value the Keplerian orbit with
\begin{equation}
\Omega_{test}=\Omega=\sqrt{\frac{GM^{(0)}(a)}{a^3}}
\end{equation}
Integrate the coupled equations
\begin{equation}
\begin{cases}
\frac{dp_0^*(R)}{dR}=\frac{2}{3}\Omega^2R-\frac{GM^{(2)}(R)}{R^2},\\
\frac{dM^{(2)}(R)}{dR}=4\pi R^2\rho(R) \frac{d\rho(R)}{dR}p_0^*(R),
\end{cases}
\end{equation}
out from the origin with boundary conditions
\begin{equation}
p_0^*(R)\rightarrow\frac{1}{3}\Omega^2R^2, \qquad M^{(2)}(R)\rightarrow0.
\end{equation}
These boundary conditions guarantee that the central density of the rotating and non-rotating configurations are the same.

\subsection{The rotating case: $l=2$ Equations}

\subsubsection{Particular Solution}

Integrate the equations
\begin{equation}
\begin{cases}
\frac{d\chi(R)}{dR}=-\frac{2GM(R)}{R^2}\varphi(R)+\frac{8\pi}{3}\Omega^2R^3 G \rho(R)\\ \nonumber
\frac{d\varphi(R)}{dR}=\left(\frac{4\pi R^2\rho(R)}{M(R)}-\frac{2}{R}\right)\varphi(R)-\frac{2\chi(R)}{GM(R)}+\frac{4\pi}{3M(R)}\rho \Omega^2 R^4
\end{cases}
\end{equation}
outward from the center with arbitrary initial conditions satisfying the equations, as $R\rightarrow 0$
\begin{eqnarray}
\varphi(R)\rightarrow A R^2, \quad \chi(R)\rightarrow B R^4, \quad B+\frac{2\pi}{3}G\rho_{c}A=\frac{2\pi}{3}G\rho_{c}\Omega^2,
\end{eqnarray}
where $A$ and $B$ are arbitrary constants. Set, for instance, $A=1$ and define $B$ from the above algebraic equation. This determines a particular solution 
$\varphi_p(R)$ and $\chi_p(R)$.

\subsubsection{Homogeneous Solution}

Integrate the homogeneous equations
\begin{equation}
\begin{cases}
\frac{d\chi_h(R)}{dR}=-\frac{2GM(R)}{R^2}\varphi_h(R)\\ \nonumber
\frac{d\varphi_h(R)}{dR}=\left(\frac{4\pi R^2\rho(R)}{M(R)}-\frac{2}{R}\right)\varphi_h(R)-\frac{2\chi_h(R)}{GM(R)}\end{cases}
\end{equation}
outward from the center with arbitrary initial conditions satisfying the equations, as $R\rightarrow 0$
\begin{eqnarray}
\varphi_h(R)\rightarrow A R^2, \quad \chi_h(R)\rightarrow B R^4, \quad B+\frac{2\pi}{3}G\rho_{c}A=0
\end{eqnarray}
This determines a particular solution $\varphi_h(R)$ and $\chi_h(R)$. Thus, the  interior solution is
\begin{eqnarray}\label{InterSol2}
\varphi_{in}(R)=\varphi_{p}(R)+K_2\varphi_{h}(R), \quad \chi_{in}(R)=\chi_{p}(R)+K_2\chi_{h}(R)
\end{eqnarray}

\subsubsection{Matching with an Exterior Solution}

The exterior solution is given as 
\begin{eqnarray}\label{ExterSol2}
\varphi_{ex}(R)&=&\frac{K_1}{R^3},\qquad \chi_{ex}(R)=\frac{K_1 G M^{(0)}}{2R^4}.
\end{eqnarray}
By matching (\ref{ExterSol2}) and (\ref{InterSol2}) at $R=a$, 
\begin{eqnarray}
\varphi_{ex}(R=a)=\varphi_{in}(R=a),\qquad \chi_{ex}(R=a)=\chi_{in}(R=a)\ ,
\end{eqnarray}
the constants $K_1$ and $K_2$ can be obtained.

The surface of the rotating configuration is described by the the polar $r_{p}$ and equatorial $r_{e}$ radii that are determined from the relationships
\begin{eqnarray}
r(a,\Theta)&=&a+\xi_{0}(a)+\xi_{2}(a)P_2(\cos\Theta),\\
r_{p}&=&r(a,0)=a+\xi_{0}(a)+\xi_{2}(a),\\
r_{e}&=&r(a,\pi/2)=a+\xi_{0}(a)-\xi_{2}(a)/2 \ .
\end{eqnarray}
In addition, the eccentricity is defined as
\begin{eqnarray}
{\rm eccentricity}=\sqrt{1-\frac{r_{p}^2}{r_{e}^2}}\ 
\end{eqnarray}
and determines completely the matching surface.

Once function $\xi_{2}(R)$ is known from Eq. (\ref{eq:53}), one can easily calculate ellipticity $\epsilon(R)$,  function $\eta_2(R)$, hence the gravitational Love number $k_2$.

\section{An example: White dwarfs}
\label{sec:dwarfs}

\begin{figure}
\includegraphics[width=3in]{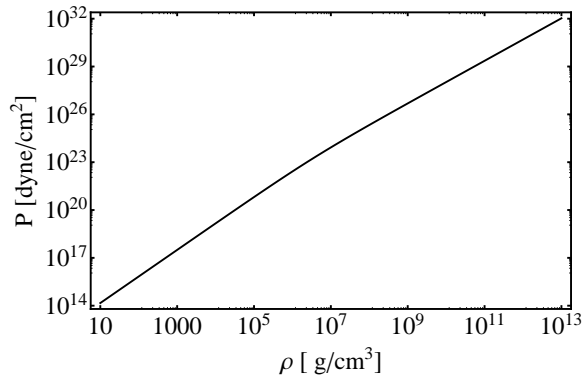} 
\caption{{\footnotesize Pressure versus density for the Chandrasekhar equation of state.
}}\label{fig:Prho} 
\end{figure}

In this section, we study an example of the formalism presented in the preceding sections to test the applicability of the method. To appreciate the validity of our results, we consider a very realistic case, namely, white dwarfs whose equation of state at zero temperature is given by the Chandrasekhar relationships \cite{rrrx}
\begin{eqnarray}
\varepsilon &=& \rho c^2=\frac{32}{3}\left(\frac{m_e}{m_n}\right)^3 K_n\left(\frac{\bar {A}}{Z}\right)x^3,\nonumber\\
p&=&\frac{4}{3}\left(\frac{m_e}{m_n}\right)^4 K_n\left[x(2x^2-3)\sqrt{1+x^2}+3\ln(x+\sqrt{1+x^2})\right] \ .
\end{eqnarray}
This means that the energy density $\varepsilon=\varepsilon(R)$ is determined by the nuclei, while the pressure $p=p(R)$ is determined by the degenerate electronic gas. Here $\bar{A}$ and $Z$ are the average atomic weight and atomic number of the corresponding nuclei; $K_n=(m_n^4c^5)/(32 \pi^2 \hbar^3)$ and $x=x(R)=p_e(R)/(m_e c)$ with $p_e(R)$, $m_e$, $m_n$, and $\hbar$ being the Fermi momentum, the mass of the electron, the mass of the nucleon and the reduced Planck constant, respectively. Here we consider the particular case  $\bar{A}/Z=2$. The behavior of the above equation of state is illustrated in Fig. \ref{fig:Prho} for the case of a degenerate electronic gas. Although the Chandrasekhar equation of state has been derived upon the basis of a phenomenological, physical approach, we see that it can be modeled with certain accuracy by means of a polytropic equation of state 
$p\propto \rho^\alpha$ with $\alpha=const.$ 

\begin{figure}
\includegraphics[width=3in]{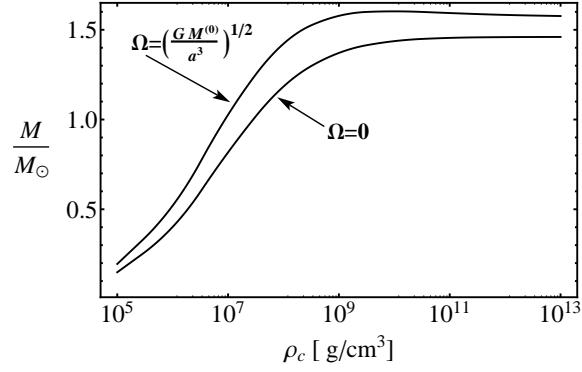} 
\caption{{\footnotesize Total mass and central density relation obtained from the Chandrasekhar equation of state.}}\label{fig:Mrho} 
\end{figure}

\begin{figure}
\includegraphics[width=3in]{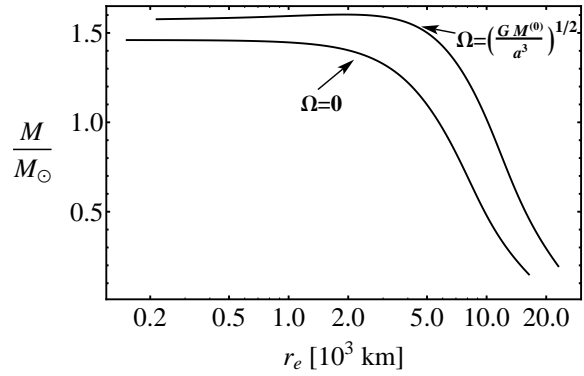} 
\caption{{\footnotesize Total mass and equatorial radius relation for Chandrasekhar equation of state.}}\label{fig:Mre} 
\end{figure}

In Fig. \ref{fig:Mrho}, we plot the behavior of the total mass as a function of the central density for a static star and for a rotating star with our test angular velocity. It is clear that for a given central density the value of the total mass is larger in the case of a rotating object than for a static body. This is in accordance with the physical expectations based upon other alternative studies \cite{psc71,jam64,ana65,rox65,chan33}. A similar behavior takes place when we explore the mass as a function of the equatorial radius, as shown in Fig. \ref{fig:Mre}.  

\begin{figure}
\includegraphics[width=3in]{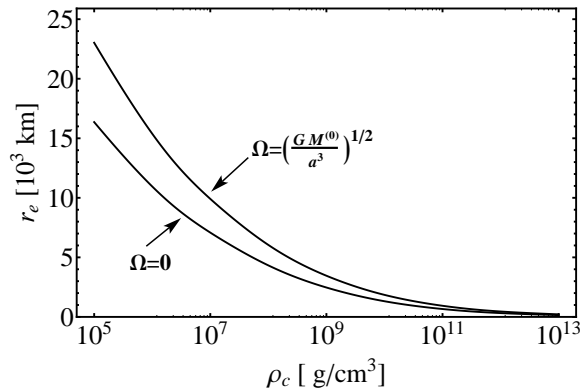} 
\caption{{\footnotesize Equatorial radius versus central density. Note that $\Omega\rightarrow 0$ as  $r_{e}\rightarrow a$.}}\label{fig:rerho} 
\end{figure}

The relationship between the central density and the equatorial radius is illustrated in Fig. \ref{fig:rerho}. As expected, the equatorial radius diminishes as the density increases, and it is larger in the case of a rotating  body. In the limit of vanishing angular velocity, the equatorial radius approaches the value of the static radius $a$. 

\begin{figure}
\includegraphics[width=3in]{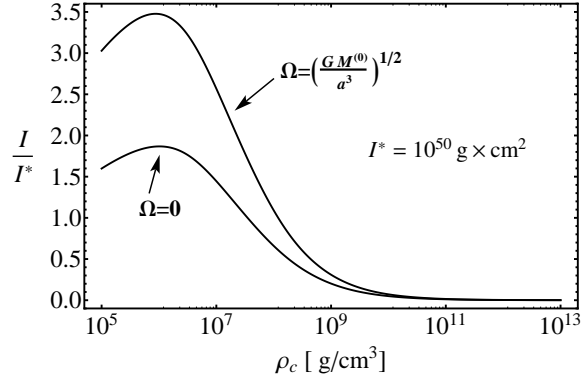} 
\caption{{\footnotesize Total moment of inertia versus central density.}}\label{fig:Mominrho} 
\end{figure}

The moment of inertia depends also on the central density and on the value of the angular velocity, as illustrated in Fig. \ref{fig:Mominrho}. For each value of the angular velocity, there is particular value of the central density at which the moment of inertia acquires a maximum. The value of the moment of inertia at the maxima increases as the angular velocity increases. For very large values of the central density, the moment of inertia turns out to be practically independent of the value of the angular velocity. Notice, however, that this happens for values close to or larger than $10^{11}$ g/cm$^3$ which should be considered as unphysical because they are larger than the critical value $\rho_c\sim 1.37 \times 10^{11}$ g/cm$^3$ at which the equation of state under consideration can no longer be applied because of the inverse $\beta$ decay process for white dwarfs consisting of helium ions. Nevertheless,  we are considering in all our plots the interval  $(10^5-10^{13})$ g/cm$^3$ for the sake of generality.

\begin{figure}
\includegraphics[width=3in]{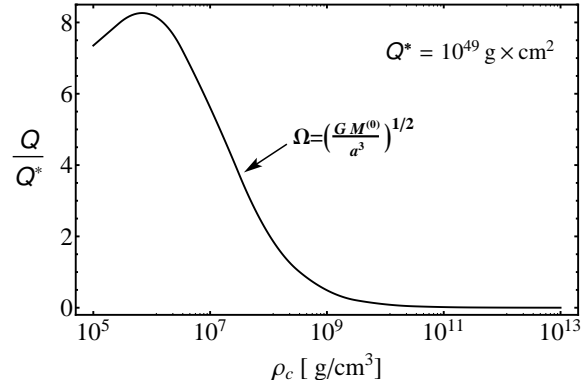} 
\caption{{\footnotesize Mass quadrupole moment versus central density.}}\label{fig:Qrho} 
\end{figure}

\begin{figure}
\includegraphics[width=3in]{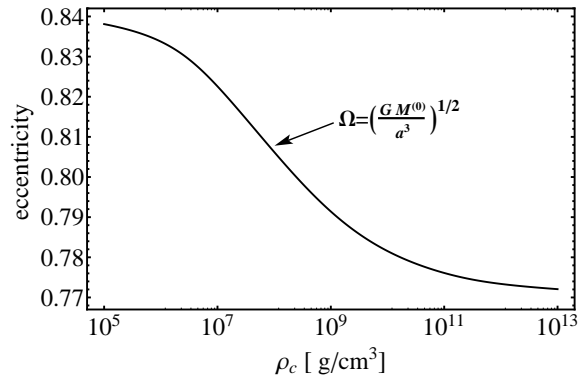} 
\caption{{\footnotesize Eccentricity versus central density of rotating configurations.}}\label{fig:eccrho} 
\end{figure}

In Figs. \ref{fig:Qrho} and \ref{fig:eccrho} 
we plot the quantities which determine the shape of the surface where the interior solution is matched with the exterior one, namely, the quadrupole moment and the eccentricity. Obviously, both quantities vanish in the limiting case of vanishing rotation. The quadrupole possesses a maximum at a certain value of the central density which coincides with the position of the maximum of the moment of inertia. 

\begin{figure}
\includegraphics[width=3in]{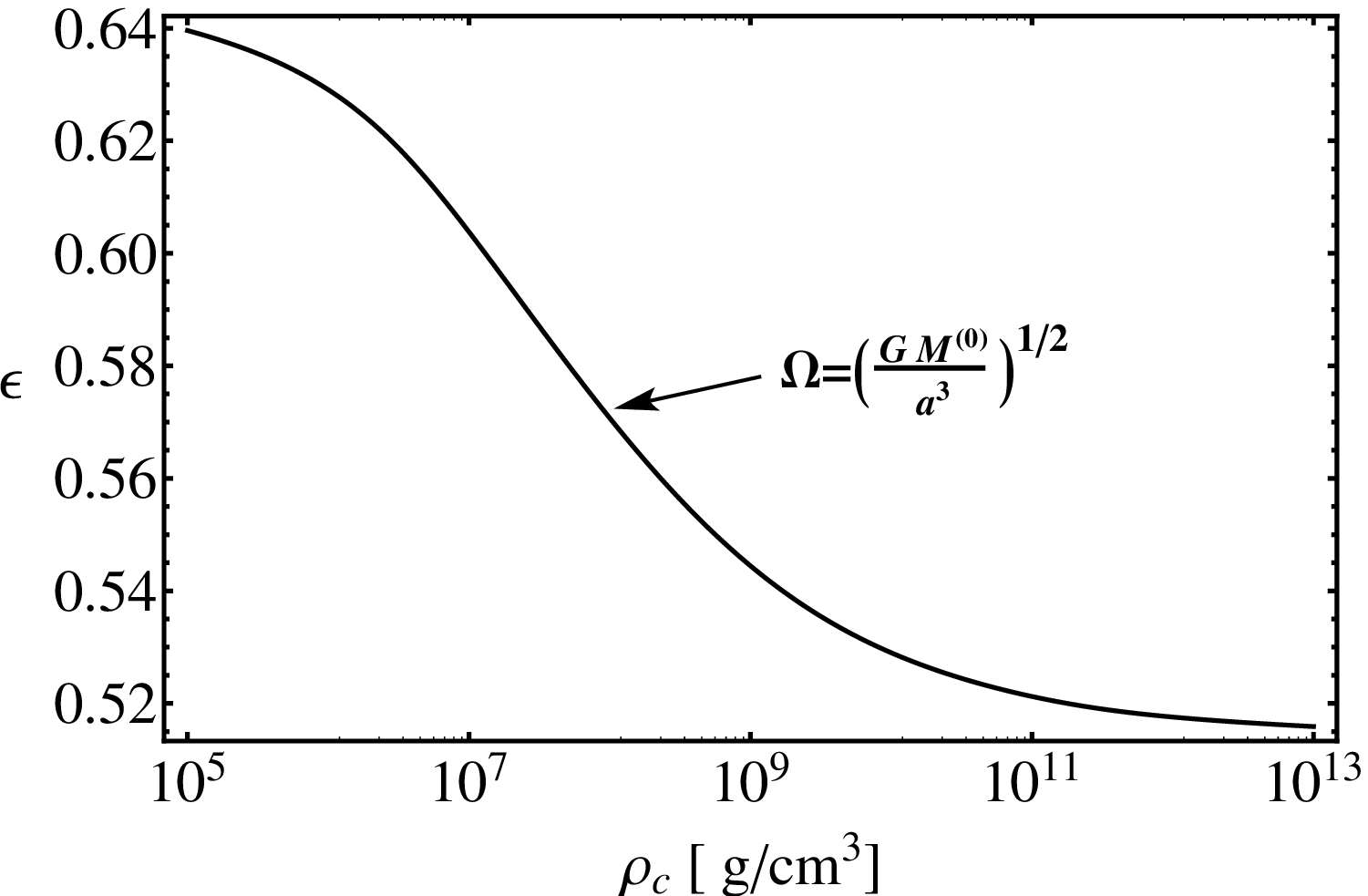} 
\caption{{\footnotesize Ellipticity versus central density of rotating configurations.}}\label{fig:elliprho} 
\end{figure}

The ellipticity of the rotating deformed star is illustrated as a function of the central density in Fig. \ref{fig:elliprho}. On the surface of the star the ellipticity shows similar behavior as the eccentricity and as density increases it decreases. Thus, the star becomes more compact and more spherical.

\begin{figure}
\includegraphics[width=3in]{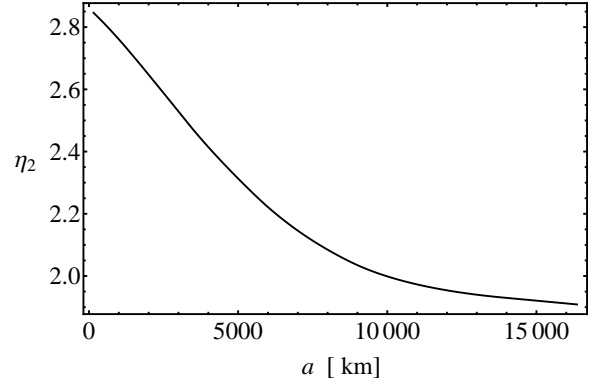} 
\caption{{\footnotesize Function $\eta_{2}$ versus spherical radius $a$ of static configurations.}}\label{fig:eta2a} 
\end{figure}

\begin{figure}
\includegraphics[width=3in]{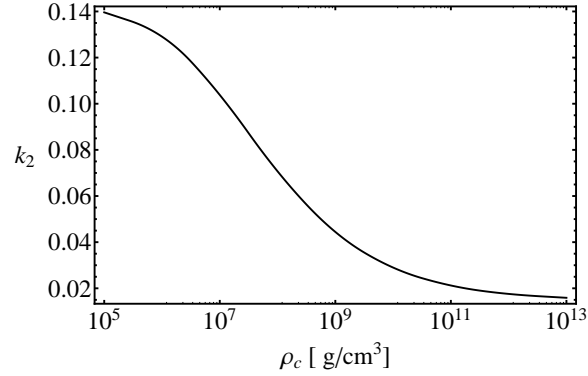} 
\caption{{\footnotesize Love number versus central density of static configurations.}}\label{fig:k2rho} 
\end{figure}

\begin{figure}
\includegraphics[width=3in]{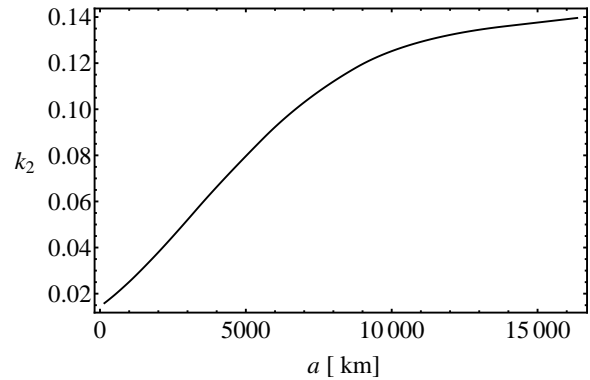} 
\caption{{\footnotesize Love number versus spherical radius of static configurations.}}\label{fig:k2a} 
\end{figure}

The dependence of function $\eta_2$ is shown as a function of the spherical radius $a$ in Fig. \ref{fig:eta2a}. As the radius increases the function decreases.The function $\eta_2$ is necessary to calculate the Love number. Finally, in Figs. \ref{fig:k2rho} and \ref{fig:k2a} we depict the gravitational Love number as a function of the central density and spherical radius, respectively. For increasing central density the Love number decreases. This implies that with the increasing central density or decreasing radius white dwarfs become less susceptible to  rotational and tidal deformations, since  $k_2=0$ for a rigid body. It should be mentioned that the values for the Love number in agreement with those presented in Ref. \cite{prodan2012}.

Notice that in all the plots, the selected values for the central mass and the equatorial radius are in accordance with the expected values for white dwarfs. We conclude that the results obtained from the numerical integration of the differential equations derived by using the approach proposed in this work are consistent with the physical expectations, when restricted to the region in which the formalism can be applied \cite{psc71, jam64, ana65, rox65, boshijmpe, boshizzo, rueda2013}.


\section{The mass-shedding limit and scaling law}
\label{kepl}
In this section we will discuss about some technical details related to the computation of the Keplerian mass-shedding limit of any rotating configurations and the scaling law for physical quantities that can be rescaled without additional numerical integrations for various objectives.

{\it The mass-shedding limit}. It is well known that the velocity of particles on the equator of the star cannot exceed the Keplerian velocity of free particles, computed at the same location. At this limit, particles on the star's surface remain bound to the star only because of a balance between gravitational and centrifugal forces. The evolution of a star rotating at the Keplerian rate is accompanied by a loss of mass, thus becoming unstable. The Keplerian angular velocity in Newtonian physics is determined as follows
\begin{equation}\label{eq:omegaK}
\Omega_{Kep}=\sqrt{\frac{GM_{tot}}{r_e^3}}
\end{equation}
where $G$ is the gravitational constant, $M_{tot}$ is the total mass of the rotating configuration and $r_e$ is the equatorial radius. This is the critical angular velocity at which rotational shedding will occur, and it is thus an upper bound on those angular velocities for which the assumption of slow rotation could be valid.

In order to estimate this quantity correctly, one needs to select a test value of the angular velocity, for example, in our computations we used $\Omega_{test}=\sqrt{GM^{(0)}/a^3}$. Usually $\Omega_{test}>\Omega_{Kep}$, hence one needs to decrease  the values of $\Omega_{test}$ gradually and estimate $\Omega_{Kep}$ successively, until $\Omega_{test}=\Omega_{Kep}$ with a given precision. For that purpose, in practice, it is convenient to use the shooting method. 

If we express $\Omega_{Kep}= \lambda \Omega_{test}$, then the value of multiplicative factor $\lambda$ can be estimated from the above procedure and the results are shown in Table~\ref{table1}.

\begin{table}[ht]
\centering
\caption{The values of multiplicative factor $\lambda$ for different values of the central density.}
\begin{center}\label{table1}
\begin{tabular}{|c|c|c|c|c|c|c|c|c|c|c|c|c|c|c|}\hline
$\rho$ [g/cm$^3$]& 10$^5$& 10$^6$& 10$^7$& 10$^8$& 10$^9$& 10$^{10}$& 5$\times$10$^{10}$& 10$^{11}$& 1.37$\times$10$^{11}$& 10$^{12}$& 10$^{13}$ \\ \hline
$\lambda$ & 0.781& 0.780& 0.777& 0.770& 0.762& 0.755& 0.752& 0.751& 0.750& 0.748& 0.747 \\ \hline
\end{tabular}
\end{center}
\end{table}

As one can see from the Table, indeed $\Omega_{test}>\Omega_{Kep}$ and this results are in agreement with the ones in the literature \cite{psc71, jam64, ana65, rox65}. It should be stressed that the Keplerian angular velocity allows one to estimate the maximum rotation rate (the minimum rotation period) and the maximum rotating mass of stars. Moreover it allows us to determine the stability region of a rotating star, inside which all rotating configurations can exist (see \cite{2013ApJ...762..117B} for details).

{\it Scaling law}. The scaling procedure is used in order to rescale all the known values of physical quantities for different objectives. For instance, the angular momentum $J$ is directly proportional to $\Omega$, hence there is always the possibility for the following scaling law $J_{new}/\Omega_{new}=J_{old}/\Omega_{old}$ to be held. This means that knowing the old value of the angular momentum $J_{old}$ for the given angular velocity $\Omega_{old}$, one can easily evaluate a new value of the angular momentum $J_{new}$ for a given new angular velocity $\Omega_{new}$ without reintegrating the structure equations. The same is true for all the physical quantities which are directly proportional to the second order of the angular velocity $\Omega^2$. The following quantities are subject to scaling:
$M^{(2)}_{new}/\Omega_{new}^2=M^{(2)}_{old}/\Omega_{old}^2$, $Q_{new}/\Omega_{new}^2=Q_{old}/\Omega_{old}^2$, $I^{(2)}_{new}/\Omega_{new}^2=I^{(2)}_{old}/\Omega_{old}^2$. From a practical point of view it is very convenient to make use of the scaling law for various computational goals.


\section{Conclusions}
\label{sec:con}

In the present work, we have revisited the Hartle formalism to describe in Newtonian gravity the structure of rotating compact objects under the condition of hydrostatic equilibrium. We use a particular set of polar coordinates that is especially constructed to take into account the deformation of the source under rotation. Moreover, we use an expansion in terms of spherical harmonics and consider all the equations only up to the second order in the angular velocity. The main point is that these assumptions allow us to reduce the problem  to a system of ordinary differential equations, instead of  partial differential equations. As a consequence, we derive all the equations explicitly and show how to perform their numerical integration.  Numerical solutions for particular equations of state and the analysis of the stability of the resulting configurations will be discussed in a subsequent work.

In addition, the formalism developed here allows us to find explicit expressions for the main physical quantities that determine the properties of the rotating configuration. In particular, we derived the equation which determines the relation between mass and central density, and showed that it  takes the form of an equation of hydrostatic equilibrium. It enforces the balance of pressure, gravitational, and centrifugal forces correctly to order $\Omega^2$. In this approximation,  the surfaces of constant density are spheroids whose ellipticity varies from zero at the center of the star up to the values  which describe the shape of the star at the surface. The ellipticity, as a function of the radius, turns out to be determined by the Clairaut's differential equation.
The equations which determine the relation between mass and central density and those which determine the ellipticity are systems of ordinary differential equations whose solution may be obtained by numerical integration. Furthermore, we also derived  analytic expressions for the quadrupole moment and moment of inertia of the source. Finally, we obtained the Clairaut-Radau equation from the Clairaut equation and calculated the gravitational Love number, which indicates rotational or tidal response to the exterior field.

We have tested the formalism developed here by using the Chandrasekhar equation of state for white dwarfs. All the derived physical quantities are in accordance with the results in the literature. This result reinforces the validity of the assumptions and approximations applied in this work to formulate a method that takes into account the rotation in the context of hydrostatic equilibrium in Newtonian gravity. Eventually, on top of everything the procedure of computing the Keplerian mass-shedding angular velocity along with the scaling law of physical quantities have been presented for rotating configurations. In view of a recent work \cite{yagi2013} on the so-called I-Love-Q relations in neutron stars and quark stars, it would be interesting to investigate these relations in white dwarf stars.

\section*{Acknowledgments}
This work was supported in part by DGAPA-UNAM, grant No. 113514, and Conacyt, grant No. 166391. 
K.B. acknowledges the support of the grants No. 3101/GF4 IPC-11, No. F.0679 and the grant for the best teachers-2015 of the Ministry of Education and Science of the Republic of Kazakhstan.

\section*{Appendix}

\appendix

\subsection*{Coordinate transformation for the Newtonian gravitational potential}

In this appendix, we show explicitly that the Newtonian field equations for $\Phi^{(0)}(r)$ in ($R, \Theta$) coordinates has the form
\begin{equation}
\nabla^2_{r}\Phi^{(0)}(r)\approx\nabla^2_{R}\Phi^{(0)}(R)+\xi(R,\Theta)\frac{d}{dR}\nabla^2_{R}\Phi^{(0)}(R)
\end{equation}
The details of the computation are as follows:
\begin{eqnarray}\nonumber
\nabla^2_{r}\Phi^{(0)}(r)&=&\left[\frac{d^2}{dr^2}+\frac{2}{r}\frac{d}{dr}\right]\Phi^{(0)}(r)=\left[\frac{dR}{dr}\frac{d}{dR}\frac{dR}{dr}\frac{d}{dR}+\frac{2}{R+\xi}\frac{dR}{dr}\frac{d}{dR}\right]\Phi^{(0)}(R+\xi)  \nonumber\\
&\approx&\left[ \left(1-\frac{d\xi}{dR}\right)\frac{d}{dR}\left(1-\frac{d\xi}{dR}\right)\frac{d}{dR}+ \frac{2}{R}\left(1-\frac{\xi}{R}\right)\left(1-\frac{d\xi}{dR}\right)\frac{d}{dR}\right]\left[\Phi^{(0)}(R)+\xi\frac{d\Phi^{(0)}(R)}{dR}\right] \nonumber\\
&\approx&\left[ \left(1-\frac{d\xi}{dR}\right) \left\{-\frac{d^2\xi}{dR^2}\frac{d}{dR}+\left(1-\frac{d\xi}{dR}\right)\frac{d^2}{dR^2}\right\}+\frac{2}{R}\left(1-\frac{\xi}{R}-\frac{d\xi}{dR}\right)\frac{d}{dR}\right]\left[\Phi^{(0)}(R)+\xi\frac{d\Phi^{(0)}(R)}{dR}\right] \nonumber\\
&\approx&\left[\frac{d^2}{dR^2}+\frac{2}{R}\frac{d}{dR}-\frac{2d\xi}{dR}\frac{d^2}{dR^2}-\frac{2}{R}\left(\frac{\xi}{R}+\frac{d\xi}{dR}+\frac{R}{2}\frac{d^2\xi}{dR^2}\right)\frac{d}{dR}\right] \left[\Phi^{(0)}(R)+\xi\frac{d\Phi^{(0)}(R)}{dR}\right] \nonumber\\
&\approx&\left[\frac{d^2}{dR^2}+\frac{2}{R}\frac{d}{dR}-\frac{2d\xi}{dR}\frac{d^2}{dR^2}-\frac{2}{R}\left(\frac{\xi}{R}+\frac{d\xi}{dR}+\frac{R}{2}\frac{d^2\xi}{dR^2}\right)\frac{d}{dR}\right]\Phi^{(0)}(R) \nonumber\\
&&+\left[\frac{d^2}{dR^2}+\frac{2}{R}\frac{d}{dR}\right]\xi\frac{d\Phi^{(0)}(R)}{dR} \nonumber\\
&=&\left[\frac{d^2}{dR^2}+\frac{2}{R}\frac{d}{dR}-\frac{2d\xi}{dR}\frac{d^2}{dR^2}-\frac{2}{R}\left(\frac{\xi}{R}+\frac{d\xi}{dR}+\frac{R}{2}\frac{d^2\xi}{dR^2}\right)\frac{d}{dR}\right]\Phi^{(0)}(R) \nonumber\\
&&+\frac{d\Phi^{(0)}(R)}{dR}\left[\frac{d^2}{dR^2}+\frac{2}{R}\frac{d}{dR}\right]\xi+\xi\left[\frac{d^2}{dR^2}+\frac{2}{R}\frac{d}{dR}\right]\frac{d\Phi^{(0)}(R)}{dR}+\frac{2d\xi}{dR}\frac{d^2\Phi^{(0)}(R)}{dR^2} \nonumber\\
&=&\left[\frac{d^2}{dR^2}+\frac{2}{R}\frac{d}{dR}-\frac{2\xi}{R^2}\frac{d}{dR}\right]\Phi^{(0)}(R)+\xi\left[\frac{d^2}{dR^2}+\frac{2}{R}\frac{d}{dR}\right]\frac{d\Phi^{(0)}(R)}{dR} \nonumber\\
&=&\left[\frac{d^2}{dR^2}+\frac{2}{R}\frac{d}{dR}\right]\Phi^{(0)}(R)+\xi\frac{d}{dR}\left[\frac{d^2}{dR^2}+\frac{2}{R}\frac{d}{dR}\right]\Phi^{(0)}(R)=\nabla^2_{R}\Phi^{(0)}(R)+\xi\frac{d}{dR}\nabla^2_{R}\Phi^{(0)}(R) \ .\nonumber
\end{eqnarray}\nonumber

\end{document}